\documentclass[twocolumn,amsmath,amssymb]{article}

\usepackage{physics}
\usepackage{ragged2e}
\usepackage{booktabs}
\usepackage{siunitx}
\usepackage{geometry}
\usepackage{xcolor}
\usepackage{subcaption}
\usepackage{times}
\usepackage{dcolumn}
\usepackage{bm}
\usepackage{graphicx}
\usepackage[superscript,biblabel]{cite}
\usepackage[mathlines]{lineno}
\usepackage{authblk}
\usepackage{cuted}
\usepackage{caption}
\usepackage[utf8]{inputenc}
\usepackage[T1]{fontenc}
\usepackage[colorlinks=true, linkcolor=blue, urlcolor=black, citecolor=blue]{hyperref}
\usepackage{sidecap}
\usepackage{stfloats}

\newcommand{\Dk}{\Delta_{\bm{k}}}
\newcommand{\gk}{\gamma(\bm{k})}
\newcommand{\Rk}{R(\bm{k})}

\usepackage{titlesec}

\titleformat{\section}
{\normalfont\small\bfseries}
{\thesection}{1em}{}

\titleformat{\subsection}
{\normalfont\footnotesize\bfseries}
{\thesubsection}{1em}{}

\usepackage{fancyhdr}
\title{Vacancy-driven inverse Lieb geometry: A general route to $d$-wave altermagnetism in two dimensions}

\footnotetext{* sasmita.mohakud@vit.ac.in}

\author{Geethanjali S$^{1}$, Katsunori Wakabayashi$^{2,}$$^{3,}$$^{4}$ and Sasmita Mohakud$^{1}$\thanks{sasmita.mohakud@vit.ac.in}}
\affil{$^{1}$ Department of Physics, School of Advanced Sciences,\\
	Vellore Institute of Technology, Vellore -- 632014, Tamil Nadu, India.\\
	$^{2}$ Department of Nanotechnology for Sustainable Energy, School of Science and Technology,
	Kwansei Gakuin University, Gakuen-Uegahara 1, Sanda, 669-1330, Japan.\\
	$^{3}$ Research Center for Materials Nanoarchitectonics (MANA), National Institute for Materials Science (NIMS), Namiki 1-1, Tsukuba 305-0044, Japan.\\
	$^{4}$ Center for Spintronics Research Network (CSRN), Osaka University, Toyonaka 560-8531, Japan.
	}

\begin{document}
	\begin{strip}
	 \vspace*{-2.5cm}
		\maketitle
		\noindent\rule{\textwidth}{0.5pt}
		\vspace{-0.1em}

		Vacancy-induced structural reconstruction provides a general microscopic route to $d$-wave altermagnetism in two-dimensional systems. As a concrete realization, reconstructed $\mathrm{V_2X_2}$ ($\mathrm{X}=\mathrm{S}, \mathrm{Se}$) monolayers form an inverse Lieb magnetic network in which two inequivalent edge vanadium sites, related by $C_4$ lattice rotational symmetry and carrying opposite exchange fields, yield zero net magnetization despite broken time-reversal ($\mathcal{T}$) and combined inversion--time-reversal ($\mathcal{PT}$) symmetries. Structural stability is confirmed by formation energies, phonon spectra, and \emph{ab initio} molecular dynamics simulations at room temperature. A minimal tight-binding model, incorporating anisotropic second-order hopping between the inequivalent magnetic sites mediated by a nonmagnetic corner site, produces spin splitting with a $(\cos k_x - \cos k_y)$ form factor in quantitative agreement with first-principles calculations. The resulting spin splitting is strongly anisotropic, maximized near the $X$ and $Y$ high-symmetry points and exhibiting a symmetry-enforced nodal degeneracy at $M$, consistent with a $d_{x^2-y^2}$ altermagnetic form factor confirmed by the fourfold Fermi surface pattern. These findings establish vacancy-driven reconstruction of an inverse Lieb magnetic network as a general design principle for two-dimensional $d$-wave altermagnets.
		
		\vspace{0.4em}
		\end{strip}

	Altermagnetism is a newly identified class of magnetic ordering in which electronic bands display momentum-dependent spin splitting driven by crystal symmetry--enforced spin--momentum locking, without net magnetization or relativistic spin--orbit coupling~\cite{vsmejkal2022emerging, mazin2022altermagnetism, bai2024altermagnetism, hayami2019momentum, cheong2025altermagnetism}. Unlike conventional antiferromagnets, where opposite-spin sublattices are related by lattice translations or spatial inversion~\cite{hayami2020bottom}, altermagnets connect them via proper or improper rotational symmetries~\cite{wei2024crystal, 9wcm-pmr2}. This distinction breaks time-reversal symmetry $\mathcal{T}$ and the combined antiunitary symmetry $\mathcal{PT}$, producing symmetry-enforced spin splitting that varies across momentum space while preserving complete spin compensation in real space. The resulting spin-split bands enable unconventional transport and optical responses, including the anomalous Hall effect~\cite{vsmejkal2020crystal, vsmejkal2022anomalous}, magneto-optical Kerr effect~\cite{samanta2020crystal, zhou2021crystal}, efficient spin-transfer torque, and giant magnetoresistance~\cite{huai2004observation, miyazaki1995giant}. Experimental confirmation of pronounced anisotropic spin splitting has been achieved in $\mathrm{MnTe}$ and $\mathrm{CrSb}$ through angle-resolved photoemission spectroscopy~\cite{lee2024broken, krempasky2024altermagnetic, osumi2024observation, reimers2024direct}, and anisotropic magnetoresistance measurements further confirm the intrinsic anisotropic spin transport~\cite{gonzalez2024anisotropic}. Altermagnets thus uniquely blend ultrafast terahertz spin dynamics with robust resistance to stray fields, positioning them as ideal platforms for next-generation spintronics~\cite{vsmejkal2022emerging}.

	While most experimentally established altermagnets are three-dimensional~\cite{bai2024altermagnetism}, two-dimensional (2D) systems offer distinct advantages: reduced thickness and enhanced structural flexibility~\cite{novoselov20162d} enable precise symmetry engineering crucial for generating momentum-dependent spin splitting. A growing number of 2D altermagnetic candidates have been identified, including $\mathrm{V_2X_2O}$ ($\mathrm{X}=\mathrm{Se}, \mathrm{Te}$)~\cite{cui2023giant}, $\mathrm{RuF_4}$~\cite{sodequist2024two, milivojevic2024interplay}, $\mathrm{MnTeMoO_6}$~\cite{zeng2024description}, and $\mathrm{Cr_2O_2}$ along with its Janus variant $\mathrm{Cr_2SeO}$~\cite{khan2025altermagnetism}. Notably, bulk $\mathrm{Mn_5Si_3}$ displays conventional antiferromagnetism yet transforms into an altermagnetic phase upon exfoliation to the 2D limit~\cite{reichlova2024observation, leiviska2024anisotropy}, underscoring the role of dimensionality in symmetry engineering. External perturbations such as strain~\cite{chakraborty2024strain} and atomic rearrangement~\cite{li2025achieving, mazin2023induced} can further tune crystal symmetries to induce altermagnetic phases. However, the microscopic origin of the characteristic $d$-wave form factor $(\cos k_x - \cos k_y)$ — and a systematic design principle for realizing it in 2D — remain elusive, leaving a fundamental gap between material predictions and mechanistic understanding.

\begin{figure*}[t]
		\centering
	    \includegraphics[width=\linewidth]{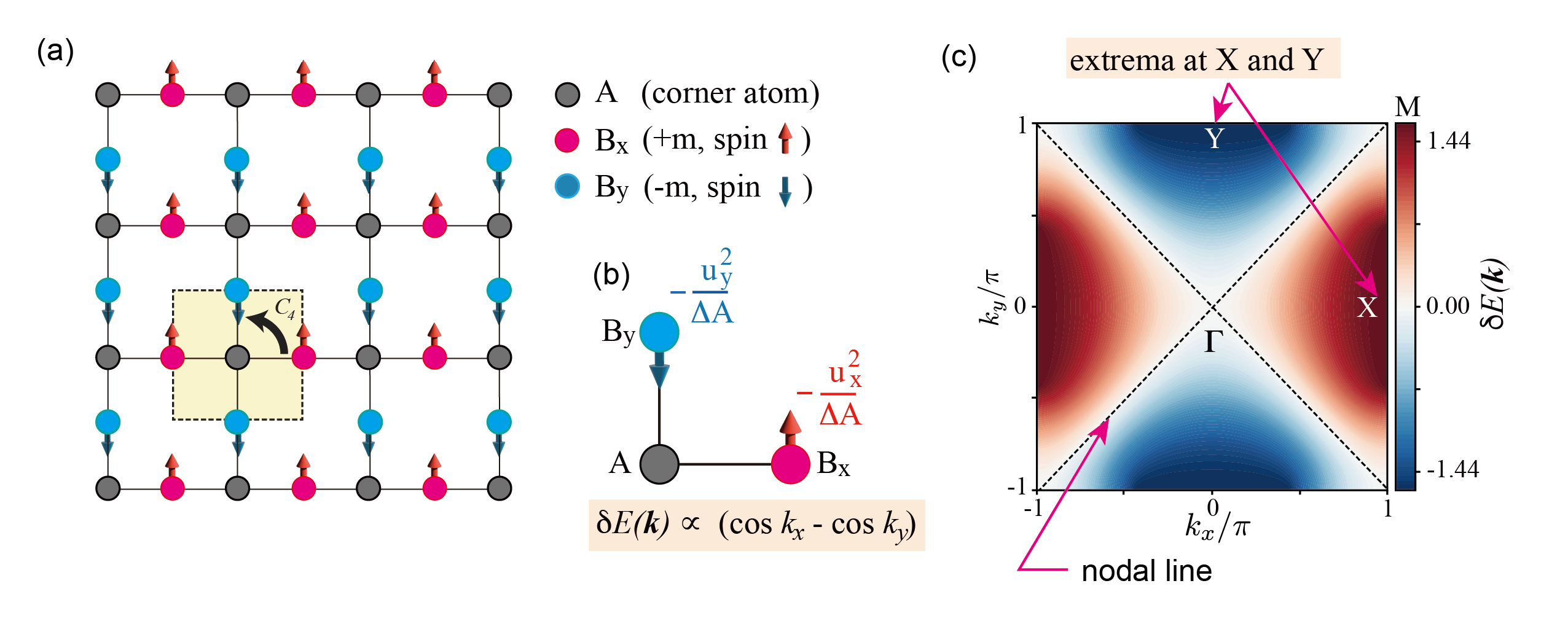}
	    \caption{
		\textbf{General mechanism for $d$-wave altermagnetism on an inverse Lieb lattice.}
		(a) Real-space inverse Lieb geometry with two magnetic edge sublattices $B_x$ and $B_y$, related by $C_4$ rotation symmetry and carrying opposite exchange fields $\pm m$, while the corner site $A$ remains nonmagnetic.
		(b) Second-order virtual hopping processes mediated by the high-energy $A$ site generate anisotropic self-energy corrections $-u_x^2/\Delta_A$ and $-u_y^2/\Delta_A$, whose difference yields the $d_{x^2-y^2}$ form factor $\alpha(\cos k_x - \cos k_y)$.
		(c) Corresponding momentum-dependent spin splitting $\delta E_{+}(\mathbf{k})$, showing a characteristic fourfold pattern with maximal splitting near the $X$ and $Y$ points and symmetry-enforced nodal lines along $k_x = \pm k_y$.
	}
	\label{fig:mechanism}
\rule{\textwidth}{0.6pt}
\end{figure*}

	In this work, we show that vacancy-induced structural reconstruction provides a general route to $d$-wave altermagnetism in two-dimensional systems. The central mechanism is the emergence of an inverse-Lieb-type magnetic network, in which two inequivalent magnetic sublattices are related by $C_4$ rotation rather than lattice translation, while anisotropic hopping mediated by a nonmagnetic intermediate site naturally generates a $\cos k_x - \cos k_y$ form factor (see Fig.~\ref{fig:mechanism}). As a concrete realization of this principle, we investigate reconstructed $\mathrm{V_2X_2}$ ($\mathrm{X}=\mathrm{S}, \mathrm{Se}$) monolayers obtained from trigonal $\mathrm{VX_2}$ by chalcogen-cluster vacancy engineering. Using first-principles calculations together with a minimal tight-binding model, we demonstrate that these systems host robust $d$-wave altermagnetic spin splitting with a characteristic nodal structure in momentum space. This combined material realization and microscopic mechanism establish vacancy-driven reconstruction as a general design principle for two-dimensional $d$-wave altermagnets.

\section*{Results and Discussion}

\subsection*{Vacancy-induced reconstruction and inverse Lieb geometry}

\begin{figure*}[t]
\centering
\includegraphics[width=0.8\linewidth]{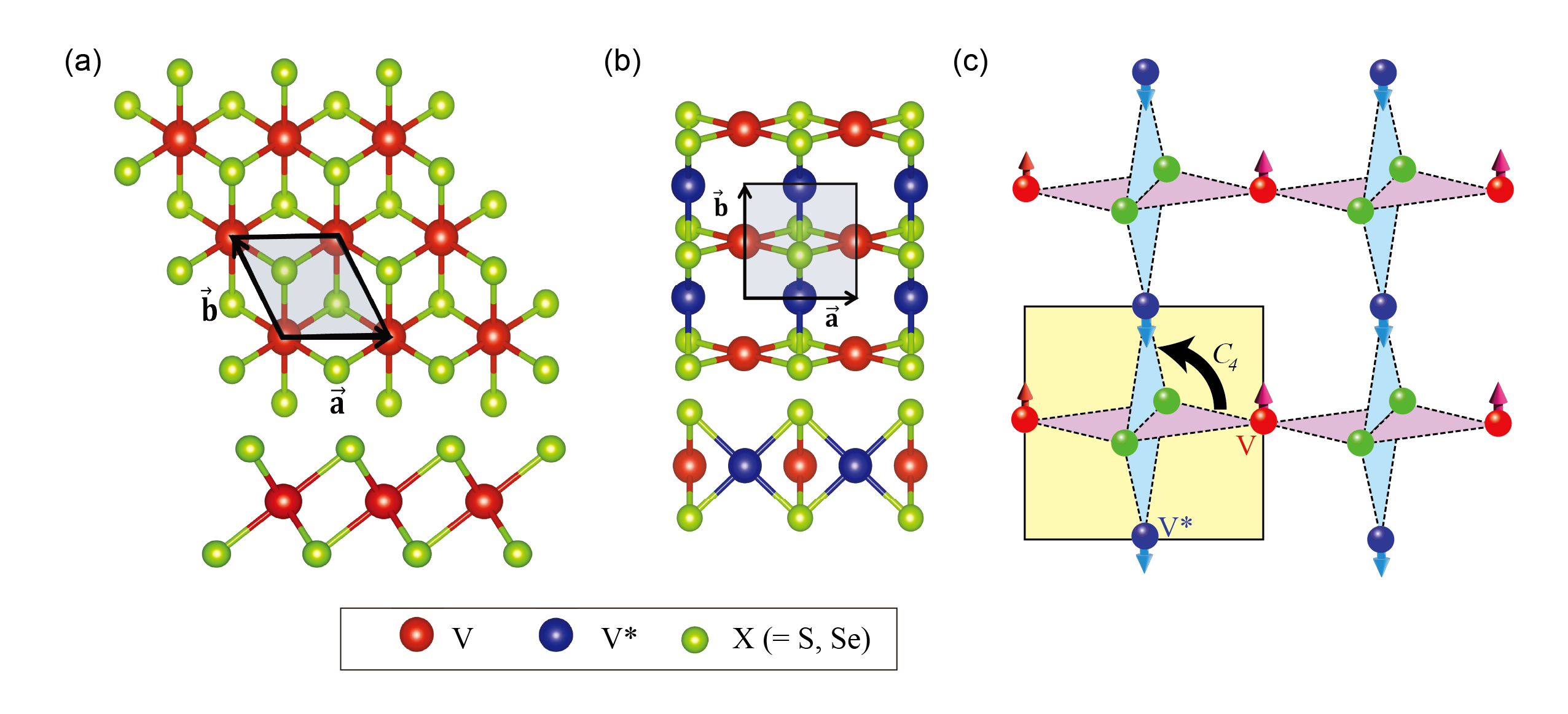}
	\caption{\textbf{Crystal structures of pristine and reconstructed materials.}
	(a) Top (upper panel) and side (lower panel) views of pristine $\mathrm{VX_2}$.
	(b) Top (upper panel) and side (lower panel) views of reconstructed monoclinic $\mathrm{V_2X_2}$, stabilized by a chalcogen-cluster vacancy. Shaded region denotes the unit cell with lattice vectors shown as arrows.
	(c) Inverse Lieb lattice geometry of the reconstructed $\mathrm{V_2X_2}$ structure, showing inequivalent vanadium sites $\mathrm{V}$ and $\mathrm{V}^*$ related by $C_4$ rotational symmetry. The intermediate sites mediate anisotropic hopping between the magnetic sublattices, forming the basis of the minimal tight-binding model.}
	\label{fig:am_change}
	\rule{\textwidth}{0.6pt}
\end{figure*}

We consider the two-dimensional trigonal crystal structure of vanadium dichalcogenides, $\mathrm{VX_2}$ ($\mathrm{X}=\mathrm{S}, \mathrm{Se}$)~\cite{li2014versatile, ma2012evidence}, shown in Fig.~\ref{fig:am_change}(a). The pristine monolayer features an X--V--X sandwich configuration, with each vanadium atom at the center of an octahedron formed by six chalcogen atoms. This octahedral coordination stabilizes the trigonal lattice symmetry and gives rise to vanadium $d$-orbital-dominated metallic bands with intrinsic ferromagnetic ordering. To engineer the electronic and magnetic properties, we introduce targeted chalcogen-cluster vacancies into the $\mathrm{VX_2}$ monolayer. Selective removal of a continuous chain of chalcogen atoms triggers extensive local bond rearrangements, forming a distinctive eight-membered atomic ring vacancy motif and changing the stoichiometry from $\mathrm{VX_2}$ to $\mathrm{V_2X_2}$. The fully optimized reconstructed $\mathrm{V_2X_2}$ structure is shown in Fig.~\ref{fig:am_change}(b).

The stability of the reconstructed $\mathrm{V_2X_2}$ ($\mathrm{X}=\mathrm{S}, \mathrm{Se}$) monolayers is confirmed through energetic, dynamical, and finite-temperature analysis. Formation energies are computed using the standard expression~\cite{ataca2012stable}:
\begin{equation}
	E_\mathrm{f}(\mathrm{V_2X_2}) = E(\mathrm{V_2X_2}) - 2E_\mathrm{V} - 2E_\mathrm{X},
\end{equation}
where $E(\mathrm{V_2X_2})$ is the total energy of the optimized $\mathrm{V_2X_2}$ supercell, and $E_\mathrm{V}$, $E_\mathrm{X}$ are the energies of vanadium and chalcogen in their stable bulk reference states. The negative values obtained for both compounds confirm energetic stability relative to the constituent elements. Phonon dispersion calculations reveal no imaginary modes across the entire Brillouin zone, establishing dynamical stability. \emph{Ab initio} molecular dynamics (AIMD) simulations at 300~K for 10~ps show total energies fluctuating tightly around their averages without structural distortion for both $\mathrm{V_2S_2}$ and $\mathrm{V_2Se_2}$, confirming thermal stability at room temperature.

Both structures adopt monoclinic symmetry (space group $P2/m$, No.~10; point group $C_{2h}$), with fully optimized lattice parameters. The vacancy-induced reconstruction generates two inequivalent vanadium sites, $\mathrm{V}$ and $\mathrm{V}^*$, each in square-planar coordination with the chalcogen atoms: $\mathrm{V}$ atoms bond to chalcogens along the $\vec{a}$ direction while $\mathrm{V}^*$ atoms bond along $\vec{b}$, creating orthogonal bonding environments that cannot be related by any lattice translation. Instead, $\mathrm{V}$ and $\mathrm{V}^*$ are connected by $C_4$ lattice rotational symmetry, and the resulting network forms an inverse Lieb lattice with magnetic vanadium atoms positioned along the edges, as shown in Fig.~\ref{fig:am_change}(c). This inverse Lieb geometry — two magnetically opposite edge sublattices related by $C_4$ rotation and coupled through a nonmagnetic corner site — is the structural prerequisite for the $d$-wave altermagnetic mechanism described below.

\subsection*{Spin-polarized electronic structure}

To elucidate the electronic and magnetic properties, we compute the spin-polarized band structure and projected density of states (PDOS) for the reconstructed $\mathrm{V_2X_2}$, presented in Fig.~\ref{fig:BANDS}(a), (b). Band crossings and a finite density of states at the Fermi level $E_\mathrm{F}$ confirm the metallic character of both compounds. A defining signature emerges along the high-symmetry points $X$ and $Y$: pronounced momentum-dependent spin splitting that alternates sign along $k_x$ and $k_y$. This behavior originates from the magnetic symmetry of the reconstructed monoclinic structure. The crystallographic point group $C_{2h} = \{E, C_2, \mathcal{P}, \sigma_h\}$ remains preserved in the magnetic phase, leaving the collinear configuration invariant. Time-reversal symmetry $\mathcal{T}$ is broken spontaneously by the finite magnetic moments on vanadium atoms, and the combined antiunitary symmetry $\mathcal{PT}$ — whose presence would enforce Kramers degeneracy via $(\mathcal{PT})^2 = -1$ — is explicitly absent in this magnetic space group. This fundamental symmetry breaking permits nonrelativistic spin splitting $E_\uparrow(\mathbf{k}) \neq E_\downarrow(\mathbf{k})$ at general $\mathbf{k}$-points, while leaving spin states degenerate at high-symmetry nodes. Analogous spin-splitting behavior has been reported for the two-dimensional altermagnet $\mathrm{RuF_4}$~\cite{milivojevic2024interplay}.

The PDOS (Fig.~\ref{fig:BANDS}(a), (b)) shows that electronic states near $E_\mathrm{F}$ are predominantly derived from the $3d$ orbitals of vanadium atoms, confirming that vanadium $3d$ electrons govern the low-energy electron and spin transport properties. Contributions from $\mathrm{S}$-$3p$ and $\mathrm{Se}$-$4p$ orbitals are negligible near $E_\mathrm{F}$. Crucially, the integrated states from opposite spins of the two inequivalent vanadium sites are equal, confirming complete spin compensation and zero net magnetization — the defining hallmark of altermagnetic order.

\subsection*{Microscopic tight-binding model}

The inverse Lieb lattice geometry of the reconstructed $\mathrm{V_2X_2}$ provides a natural basis for a minimal microscopic model. The key structural features translate directly into the essential ingredients of a three-site tight-binding model: two inequivalent vanadium edge sites with opposite exchange fields are coupled through a nonmagnetic corner site, with directional bonding along $\vec{a}$ and $\vec{b}$ introducing direction-dependent hopping amplitudes. Since the PDOS confirms that states near $E_\mathrm{F}$ are dominated by vanadium $d$ orbitals while the chalcogen-derived states are at higher energy, the corner $A$ site can be treated as a high-energy sector and integrated out, leaving an effective two-band description of the low-energy physics.

We construct the three-site model on the inverse Lieb lattice (Fig.~\ref{fig:mechanism}(a)) with one high-energy corner site ($A$) and two inequivalent edge sites ($B_x$, $B_y$), corresponding to $\mathrm{V}$ and $\mathrm{V}^*$. Defining the spinor basis
\begin{equation}
	\Psi_{\bm{k}s} =
	\bigl(c_{A,\bm{k}s},\; c_{B_x,\bm{k}s},\; c_{B_y,\bm{k}s}\bigr)^T,
	\qquad s = \uparrow(\!+\!1),\,\downarrow(\!-\!1),
\end{equation}
the $3\times 3$ spin-resolved Hamiltonian reads
\begin{equation}
	H_s(\bm{k}) =
	\begin{pmatrix}
		\epsilon_A          & u_x(\bm{k})      & u_y(\bm{k})      \\
		u_x(\bm{k})         & \epsilon_B + sm  & 0                \\
		u_y(\bm{k})         & 0                & \epsilon_B - sm
	\end{pmatrix},
	\label{eq:H3}
\end{equation}
with nearest-neighbor hopping amplitudes $u_x(\bm{k}) = 2t\cos(k_x/2)$ and $u_y(\bm{k}) = 2t\cos(k_y/2)$. The exchange field $m > 0$ acts with opposite signs on $B_x$ and $B_y$, ensuring complete spin compensation ($\langle M \rangle = 0$) while breaking time-reversal symmetry.

Treating the $A$ site as the high-energy sector ($\Delta_A \equiv \epsilon_A - \epsilon_B \gg |u_{x,y}|$) and integrating it out via L\"{o}wdin partitioning, and introducing the average dispersion $\varepsilon(\bm{k}) = \epsilon_B - (t^2/\Delta_A)(\cos k_x + \cos k_y)$, the effective $2\times 2$ Hamiltonian takes the compact Pauli-matrix form:
\begin{equation}
	H_{\mathrm{eff}}^{(s)}(\bm{k}) =
	\varepsilon(\bm{k})\,\rho_0
	+ \bigl(sm - \alpha\Dk\bigr)\rho_z
	- \beta\,\gk\,\rho_x,
	\label{eq:Heff_pauli}
\end{equation}
where $\rho_i$ are Pauli matrices in the $(B_x, B_y)$ sublattice space, $\alpha = 2t^2/\Delta_A$, $\beta = 4t^2/\Delta_A$, $\Dk = \cos k_x - \cos k_y$, and $\gk = \cos(k_x/2)\cos(k_y/2)$. The term $\alpha\Dk$ arises directly from the lattice geometry: second-order virtual hopping $B_x \to A \to B_x$ and $B_y \to A \to B_y$ generate self-energy corrections $-u_x^2/\Delta_A$ and $-u_y^2/\Delta_A$ whose difference yields the $d$-wave form factor (Fig.~\ref{fig:mechanism}(b)).

\begin{figure*}[t]
	\centering
	\includegraphics[width=\linewidth]{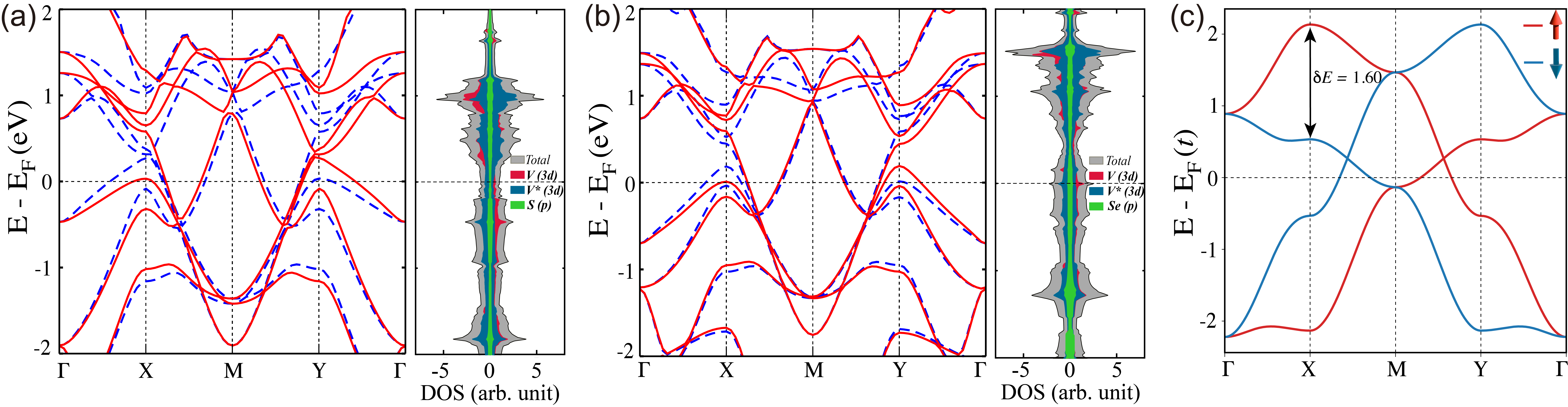}
	\caption{\textbf{Spin-polarized electronic band structure and minimal-model comparison.}
	Spin-resolved band structures (left panels) and projected density of states (PDOS, right panels) for (a) $\mathrm{V_2S_2}$ and (b) $\mathrm{V_2Se_2}$. Solid red and dashed blue lines denote spin-up and spin-down bands, respectively, computed along $\Gamma$--$X$--$M$--$Y$--$\Gamma$. Pronounced momentum-dependent spin splitting is visible near $X$ and $Y$, with degeneracy at $M$. 
	(c) Spin-resolved band structure from the minimal tight-binding model, Eq.~\eqref{eq:eigenvalues}, with parameters $t = 1$, $\Delta_A = 3$, $m = 0.8$ (in units of $t$). The model reproduces the essential features of the DFT results, including maximal splitting at $X$ and $Y$ and the $d_{x^2-y^2}$ nodal structure. The black double-headed arrow marks $\delta E_+$ at the $X$ point.}
	\label{fig:BANDS}
	\rule{\textwidth}{0.6pt}
\end{figure*}

Diagonalizing Eq.~\eqref{eq:Heff_pauli} yields the energy eigenvalues:
{\footnotesize
\begin{equation}
	E_{\pm,s}(\bm{k}) =
	\varepsilon(\bm{k}) \pm
	\sqrt{\bigl(sm - \alpha(\cos k_x - \cos k_y)\bigr)^2
		+ \beta^2\gamma^2(\bm{k})}.
	\label{eq:eigenvalues}
\end{equation}}
The spin-resolved band structure computed from Eq.~\eqref{eq:eigenvalues} along $\Gamma$--$X$--$M$--$Y$--$\Gamma$ is shown in Fig.~\ref{fig:BANDS}(c), reproducing the pronounced splitting near $X$ and $Y$ and the degeneracy at $M$ in excellent agreement with the first-principles results. The spin splitting of the $\pm$ bands,
\begin{equation}
	\delta E_\pm(\bm{k}) \equiv E_{\pm,\uparrow} - E_{\pm,\downarrow}
	\approx \mp\frac{2m\alpha\,(\cos k_x - \cos k_y)}{\Rk},
\end{equation}
with $\Rk = \sqrt{m^2 + \alpha^2\Dk^2 + \beta^2\gamma^2(\bm{k})}$, is directly proportional to $(\cos k_x - \cos k_y)$, with magnitude controlled by the exchange field $m$ and second-order hopping coefficient $\alpha = 2t^2/\Delta_A$. In the exchange-dominated limit $m \gg \alpha|\Dk|,\,\beta|\gk|$, this reduces to the pure $d_{x^2-y^2}$ form:
\begin{equation}
	\delta E_\pm(\bm{k}) \approx
	\mp\,2\alpha\,(\cos k_x - \cos k_y).
	\label{eq:dwave}
\end{equation}
The $d$-wave symmetry of the splitting pattern is preserved for all parameter values, while the magnitude is modulated by $1/\Rk$. At $\bm{k} = M = (\pi, \pi)$, both $\Dk$ and $\gk$ vanish exactly by $C_4$ symmetry, enforcing the nodal degeneracy. Along $\Gamma$--$X$ and $\Gamma$--$Y$, $\Dk$ changes sign, producing the characteristic fourfold angular modulation. The $C_4$ rotation $(k_x, k_y) \to (-k_y, k_x)$ maps $\Dk \to -\Dk$, enforcing $\delta E(\bm{k}) = -\delta E(C_4\bm{k})$ — the symmetry origin of the $d_{x^2-y^2}$ form factor. Detailed derivations are provided in the supplementary information.

\subsection*{$d_{x^2-y^2}$ symmetry and Fermi surface topology}

\begin{figure*}
	\centering
	\includegraphics[width=\linewidth]{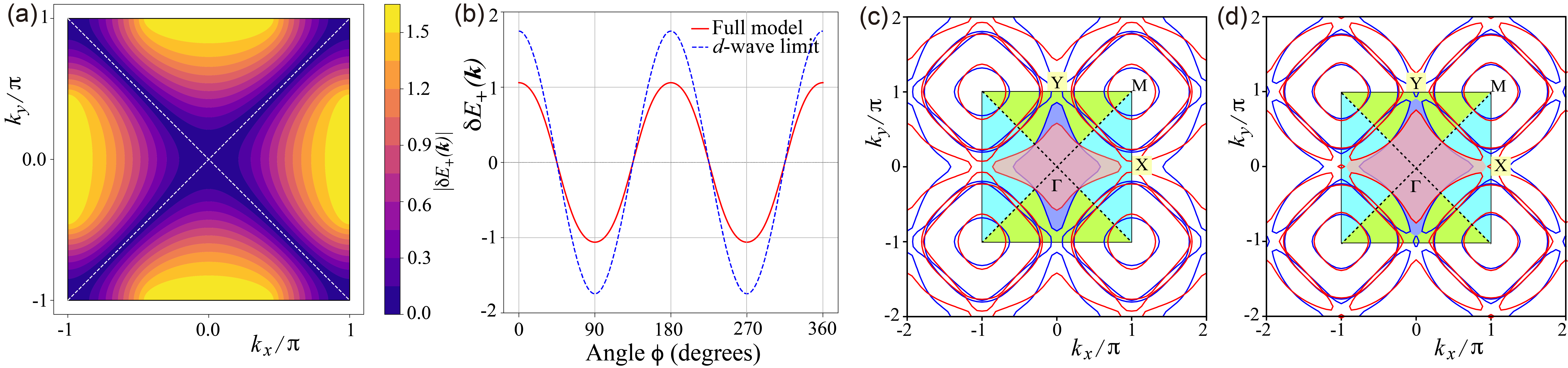}
	\caption{\textbf{$d$-wave symmetry of the altermagnetic spin splitting and spin-resolved Fermi surface.}
		(a) Magnitude $|\delta E_+(\bm{k})|$ in the first Brillouin zone. Dashed white lines indicate the symmetry-enforced nodal lines along $k_x = \pm k_y$.
		(b) Angular dependence of $\delta E_+$ on a circle of radius $|\bm{k}| = 0.6\pi$. The solid red curve is the full model result; the dashed blue curve is the clean $d$-wave approximation $\propto \cos k_x - \cos k_y$. The nearly identical profiles confirm the validity of the $d$-wave limit at this wave vector.
		Spin-resolved Fermi surface contours of (c) $\mathrm{V_2S_2}$ and (d) $\mathrm{V_2Se_2}$ in the extended Brillouin zone $(k_x, k_y) \in [-2\pi, 2\pi]$. Solid red and blue curves represent opposite spin channels, revealing momentum-dependent altermagnetic spin splitting with a characteristic fourfold sign-changing pattern (highlighted region in the first Brillouin zone) consistent with the $d_{x^2-y^2}$ form factor. Spin-degenerate nodes appear at $M$ points, while maximum splitting occurs near $X$ and $Y$.}
	\label{fig:fermi_surface}
	\rule{\linewidth}{0.6pt}
\end{figure*}

The two-dimensional map of $|\delta E_+(\bm{k})|$ and the angular dependence at fixed $|\bm{k}| = 0.6\pi$ are shown in Figs.~\ref{fig:fermi_surface}(a), (b). The magnitude is maximal near $X$ and $Y$ and vanishes along the nodal lines $k_x = \pm k_y$, while the angular profile closely matches the clean $d$-wave approximation $\propto \cos k_x - \cos k_y$, confirming the validity of the $d_{x^2-y^2}$ limit at this wave vector.

The spin-resolved Fermi surface contours of $\mathrm{V_2S_2}$ and $\mathrm{V_2Se_2}$ in the extended Brillouin zone are shown in Figs.~\ref{fig:fermi_surface}(c), (d). Spin-up (red) and spin-down (blue) contours exhibit clear momentum-dependent separation, confirming finite spin splitting at $E_\mathrm{F}$. The splitting is highly anisotropic: it is enhanced along $\Gamma$--$X$ and $\Gamma$--$Y$ and becomes degenerate at $M$, with a fourfold sign-changing pattern characteristic of the $d_{x^2-y^2}$ altermagnetic form factor. This anisotropic Fermi surface topology is a key signature of altermagnetism, where the spin polarization alternates in momentum space rather than producing a uniform ferromagnetic-like splitting. While both compounds display qualitatively similar $d$-wave-like behavior, $\mathrm{V_2S_2}$ shows a noticeably larger spin splitting and more pronounced Fermi surface wrapping, indicating a stronger effective altermagnetic interaction at $E_\mathrm{F}$. The anisotropic spin splitting and associated spin transport are accessible experimentally via angle-resolved photoemission spectroscopy (ARPES) and anisotropic magnetoresistance (AMR) measurements.

\subsection*{Real-space spin texture}

\begin{figure*}
		\centering
		\includegraphics[width=0.7\linewidth]{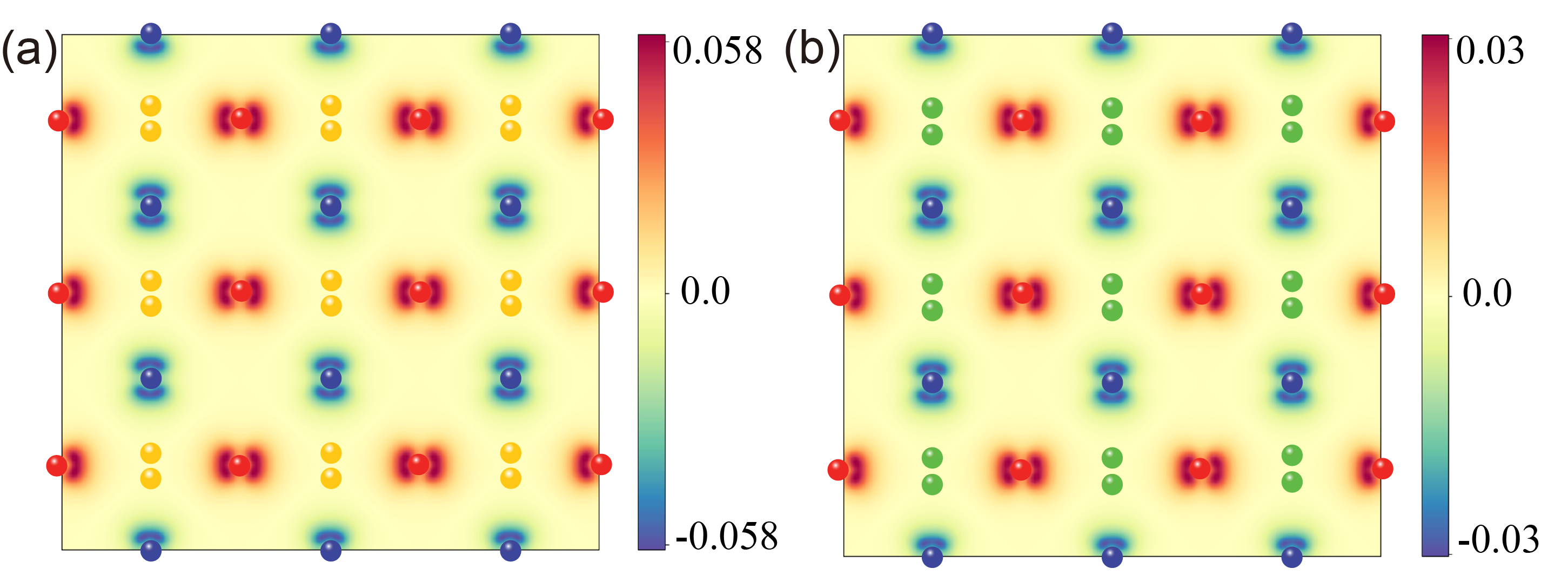}
		\caption{\textbf{Real-space spin density distribution.}
			Top views of spin-density distributions of (a) $\mathrm{V_2S_2}$ and (b) $\mathrm{V_2Se_2}$. Red and blue colors represent spin-up and spin-down densities, respectively, highlighting the antiparallel spin polarization localized on the inequivalent vanadium sublattices $\mathrm{V}$ and $\mathrm{V}^*$.}
		\label{fig:spin_density}
		\rule{\linewidth}{0.6pt}
\end{figure*}

The real-space manifestation of the altermagnetic order is captured by the spin density, computed as the difference between spin-up and spin-down densities and shown in Fig.~\ref{fig:spin_density}(a), (b). The $\mathrm{V}$ atoms predominantly host spin-up polarization while $\mathrm{V}^*$ atoms carry spin-down polarization of equal magnitude, confirming the compensated magnetic state and the localized character of the vanadium moments. The spin density around each vanadium atom displays a characteristic two-lobed anisotropic distribution consistent with the underlying $d$-orbital symmetry. This sublattice-resolved spin texture — with opposite spins related by $C_4$ rotational and $C_2$ spin symmetry rather than simple translation — constitutes the real-space signature of altermagnetic order while preserving zero net magnetization. Together, the minimal tight-binding model, the momentum-resolved Fermi surface analysis, and the real-space spin density provide a unified and self-consistent confirmation of $d$-wave altermagnetic order in the reconstructed $\mathrm{V_2X_2}$ monolayers.

\section*{Summary}

We have demonstrated that vacancy-driven lattice reconstruction in trigonal $\mathrm{VX_2}$ ($\mathrm{X}=\mathrm{S}, \mathrm{Se}$) monolayers leads to stable monoclinic $\mathrm{V_2X_2}$ structures whose inverse Lieb geometry provides a general microscopic route to $d$-wave altermagnetism in two dimensions. Structural stability is confirmed by negative formation energies, absence of imaginary phonon modes, and thermally stable \emph{ab initio} molecular dynamics at room temperature. The reconstructed geometry forms an inverse Lieb lattice in which two magnetic sublattices related by $C_4$ rotational symmetry carry opposite exchange fields, while anisotropic second-order hopping mediated by a nonmagnetic corner site naturally generates a $\cos k_x - \cos k_y$ form factor. A minimal tight-binding model analytically captures this mechanism and quantitatively reproduces the essential features of the first-principles band structure: the characteristic $d_{x^2-y^2}$ spin splitting, the symmetry-enforced nodal degeneracy at $M$, and the fourfold Fermi surface pattern. The real-space spin density confirms sublattice-localized opposite polarization with zero net magnetization, consistent with altermagnetic order.

These results establish vacancy-driven reconstruction of an inverse Lieb magnetic network as a general design principle for engineering $d$-wave altermagnetism in two-dimensional systems. The mechanism identified here is not restricted to $\mathrm{V_2X_2}$ or to chalcogenide chemistry. Any two-dimensional lattice hosting (i) two magnetic sublattices related by a proper rotational symmetry and (ii) anisotropic hopping mediated by an intermediate nonmagnetic site is expected to exhibit the same $d_{x^2-y^2}$ form factor, regardless of the specific atomic species or bonding type. Vacancy engineering provides a direct experimental route to realize this geometry in a wide class of transition-metal dichalcogenides and related layered compounds. This positions $\mathrm{V_2X_2}$ monolayers as a concrete proof-of-concept, while pointing toward a broader family of two-dimensional $d$-wave altermagnets accessible through controlled defect engineering for spintronic applications exploiting spin-selective transport and anomalous Hall effects.

\section*{Computational Details}
To investigate the structural, electronic, and magnetic properties of the system, we perform first-principles calculations within the framework of density functional theory (DFT) as implemented in the Quantum ESPRESSO package~\cite{giannozzi2009quantum}. The generalized gradient approximation (GGA) with the Perdew--Burke--Ernzerhof (PBE) functional~\cite{perdew1996generalized} is used to treat exchange-correlation effects. A plane-wave basis set with a kinetic energy cutoff of 70~Ry and a self-consistent-field convergence threshold of $10^{-8}$~Ry are employed. Brillouin zone sampling uses a $\Gamma$-centered $12\times12\times1$ $k$-point mesh with Methfessel--Paxton smearing of 0.07~Ry. Structural relaxations are performed until residual forces on each atom fall below 0.02~Ry/Bohr. A vacuum layer of 20~\AA{} along the $z$-direction eliminates interactions between periodic images. Phonon dispersions are computed using the Vienna \emph{ab initio} Simulation Package (VASP)~\cite{kresse1996efficient, kresse1999ultrasoft} via density-functional perturbation theory~\cite{togo2015first}. Thermal stability is assessed via \emph{ab initio} molecular dynamics (AIMD) simulations in VASP using a Nos\'{e}--Hoover thermostat at 300~K with 2~fs timesteps over 5000 steps (10~ps total). All calculations are performed without spin--orbit coupling (SOC). Because the altermagnetic spin splitting reported here is symmetry-enforced and nonrelativistic in origin, SOC is expected to introduce only minor quantitative corrections without altering the $d_{x^2-y^2}$ form factor or the symmetry-protected nodal structure.

\section*{Acknowledgements}
SM acknowledges the SEED grant (RGEMS) (No.\ SG20230062) funded by Vellore Institute of Technology (VIT), Vellore, India, and the Science and Engineering Research Board (SERB), Department of Science and Technology (DST), Government of India (No.\ ECR/2016/001431) for financial support. 
KW acknowledges	the financial support from	JSPS KAKENHI (Grants No.~JP25K01609, No.~JP22H05473), JST CREST (Grant
No.~JPMJCR19T1) and Basic Science Research Projects (Grant No.~2401203) from the Sumitomo Foundation. 
\bibliographystyle{unsrt}
\bibliography{references}

\end{document}